\documentclass{amsproc}

\newtheorem{theorem}{Theorem}[section]
\newtheorem{lemma}[theorem]{Lemma}

\theoremstyle{definition}

\theoremstyle{remark}

\numberwithin{equation}{section}

\begin{document}

\title[Schr\"odinger Operators with a
Surface-Supported $\delta$ Interaction]{Spectral Properties of
Schr\"odinger Operators \\ with a Strongly Attractive $\delta$
Interaction \\ Supported by a Surface}

\author{Pavel Exner}
\address{Department of Theoretical Physics, Nuclear Physics
Institute, Academy of Sciences, 25068 \v{R}e\v{z} near Prague,
Czech Republic} \email{exner@ujf.cas.cz}
\thanks{The author obtained a support from NSF for a part of
conference expenses and from GAAS Grant \#1048101.}

%    General info
\date{September ??, 2002 and, in revised form, ????, 2002.}


\begin{abstract}
We investigate the operator $-\Delta -\alpha \delta (x-\Gamma)$ in
$L^2(\mathbb{R}^3)$, where $\Gamma$ is a smooth surface which is
either compact or periodic and satisfies suitable regularity
requirements. We find an asymptotic expansion for the lower part
of the spectrum as $\alpha\to\infty$ which involves a
``two-dimensional'' comparison operator determined by the geometry
of the surface $\Gamma$. In the compact case the asymptotics
concerns negative eigenvalues, in the periodic case Floquet
eigenvalues. We also give a bandwidth estimate in the case when a
periodic $\Gamma$ decomposes into compact connected components.
Finally, we comment on analogous systems of lower dimension and
other aspects of the problem.
\end{abstract}

\maketitle

\section{Introduction}

The aim of the present paper is to discuss asymptotic spectral
properties of a class of generalized Schr\"{o}dinger operators in
$L^2(\mathbb{R}^3)$. The corresponding potential will be a
negative multiple of the Dirac measure supported by a surface
$\Gamma\subset \mathbb{R}^3$. In other words, we are going to
treat operators corresponding to the formal expression
%---------------%
\begin{equation} \label{formal}
-\Delta -\alpha \delta (x-\Gamma )\,,
\end{equation}
%---------------%
where $\alpha>0 $ is independent of $x$; properties of $\Gamma$
will be specified below.

Apart from being an interesting mathematical question in itself,
the problem has a natural motivation coming from quantum
mechanics. Long time ago, physicists considered a formal
``shrinking limit'' for a particle localized in the vicinity of a
manifold as a natural approach to quantization \cite{JK, To, dC}.
These considerations inspired studies of spectral and scattering
properties of ``fat'' curved manifolds -- see \cite{DE, DEK} and
references therein. Recently the mentioned limiting argument was
reconsidered on a rigorous footing \cite{FH}, and related results
were obtained for other geometric structures such as planar graphs
\cite{RuS, KZ}. The operator which played role of the Hamiltonian
in these models was (a multiple of) the Dirichlet Laplacian in a
neighborhood of the manifold. Sometimes other boundary conditions
were used: in the papers \cite{RuS, KZ} the Dirichlet boundary
condition was replaced by the Neumann one.

Another natural physical application, namely modeling of electron
behavior in quantum wires, semiconductor thin films, and similar
structures, motivates us to consider a modification of the above
scheme in which the particle would be less strictly coupled to the
manifold. A way to achieve this goal is to adopt the operator
(\ref{formal}) as the Hamiltonian of such a system. It is clear
that the confinement in this model takes place at negative
energies only. Moreover, such a particle can be found at large
distances from $\Gamma$, although with a small probability,
because the exterior of the manifold is a classically forbidden
region.

The ``shrinking limit'' then corresponds to making the $\delta$
coupling strong. The main idea is that for large $\alpha$ the
eigenfunctions of the operator (\ref{formal}) are localized close
to $\Gamma$. We employ a two-sided estimate of the eigenvalues
using minimax principle in combination with a bracketing argument.
We take a layer-type neighborhood of $\Gamma$ and impose Dirichlet
and Neumann conditions at its boundary. In view of the strong
localization we obtain in this way precise bounds for the negative
part of the spectrum. Using the sketched method we have been able
to get asymptotic behavior of the eigenvalues as $\alpha\to\infty$
for the operator (\ref{formal}) in $L^2(\mathbb{R}^2)$ with
$\Gamma$ belonging to various curve classes -- see \cite{EY1, EY2}
and \cite{EY3} for planar loops with a magnetic field. In
\cite{EK} the argument was extended to surfaces in $\mathbb{R}^3$
which are diffeomorphic to $\mathbb{R}^2$ and asymptotically
planar; the aim was to show that under additional assumptions the
operator (\ref{formal}) has then a nontrivial discrete
spectrum\footnote{The analogous result in the two-dimensional case
was proved under weaker assumptions in \cite{EI} and for a curve
in $\mathbb{R}^3$ in \cite{EK2}; in both cases the $\delta$
coupling is not required to be strong.}.

In the present paper we are going to treat the analogous problem
for two other classes of manifolds without a boundary. The first
are compact surfaces, the second periodic ones; we will derive an
asymptotic expansion for the eigenvalues in the former case and
for Floquet eigenvalues in the latter. An important question for
periodic systems is the existence of spectral gaps. In case of a
nontrivial periodic curve \cite{EY2} open gaps always exist for
$\alpha$ large enough. This is not true for surfaces. However, we
will be able to prove the existence of gaps for a class of
non-connected $\Gamma$. In conclusion we will comment on
extensions of the mentioned two-dimensional results and some other
aspects of the problem.

\section{Compact surfaces}

\subsection{Formulation of the problem and the results}

Let $\Gamma \subset \mathbb{R}^3$ be a $C^4$ smooth compact
Riemann surface of a finite genus $g$, i.e. diffeomorphic to a
sphere with $g$ handles attached \cite{Kli}. As such, it can be
parameterized by a finite atlas. The $i$-th chart $p_i:\,U_j \to
\mathbb{R}^3$ can be expressed in local coordinates
$s^{(i)}_\mu,\; \mu=1,2$, the particular choice of which will not
be important in the following. Usually, we will suppress the chart
index. The metric tensor given in the local coordinates by
$g_{\mu\nu }=p_{,\mu }\cdot p_{,\nu}$ defines the invariant
surface area element $d\Gamma :=g^{1/2}d^2 s$, where $g:=\det
(g_{\mu\nu})$. Furthermore, the tangent vectors $p_{,\mu}$ are
linearly independent, and their cross product $p_{,1 }\times
p_{,2}$ gives, after rescaling, a unit normal field $n$ on
$\Gamma$. The Weingarten tensor is then obtained by raising the
index in the second fundamental form, $h_{\mu}\, ^{\nu
}:=-n_{,\mu}\cdot p_{,\sigma }g^{\sigma\nu}$, where $(g^{\mu\nu})$
means conventionally $(g_{\mu\nu})^{-1}$. The eigenvalues
$k_{\pm}$ of $(h_{\mu }\,^{\nu})$ are the principal curvatures.
They determine the Gauss curvature $K$ and mean curvature $M$ by
% ------------- %
\begin{equation} \label{defiKM}
K=\det (h_{\mu }\,^{\nu })=k_{+}k_{-}\,,\quad
M=\frac{1}{2}\,\mathrm{Tr\:} (h_{\mu }\,^{\nu})
=\frac{1}{2}(k_{+}\!+k_{-})\,.
\end{equation}
% ------------- %
The object of our interest is the generalized Schr\"odinger
operator with an attractive measure-type potential. The latter is
a multiple of the Dirac measure $\mu_{\Gamma}$ defined by
$\mu_{\Gamma}(B) :=\mathrm{vol}(B\cap\Gamma)$ for any Borel
$B\in\mathbb{R}^3,$ where $\mathrm{vol}(\cdot)$ is two-dimensional
Hausdorff measure on $\Gamma$. Using the trace map $W^{2,1}
(\mathbb{R}^3)\to L^2(\mathbb{R}^3,\mu_\Gamma)\cong
L^2(\Gamma,d\Gamma)$ which is well defined in view of a standard
Sobolev embedding, and abusing slightly the notation, we can
define the quadratic form
%--------------%
\begin{equation} \label{form}
q_{\alpha}\left[ \psi \right] = \| \nabla
\psi\|^2_{L^2(\mathbb{R}^3)} -\alpha \int_{\mathbb{R}^3 }
|\psi(x)|^{2} d\mu_\Gamma(x)\,, \quad \psi \in
W^{2,1}(\mathbb{R}^3)\,.
\end{equation}
%--------------%
By Theorem~4.2 of \cite{BEKS} this form is bounded from below and
closed. Therefore, it is associated with a unique semibounded
self-adjoint operator $H_{\alpha,\Gamma}$ which is regarded as the
realization of the formal expression (\ref{formal}). Let us remark
that since $\Gamma$ is smooth one can define the operator
$H_{\alpha,\Gamma}$ alternatively through boundary conditions
which involve the jump of the normal derivative across the surface
in the same way as in \cite{EK}. This corresponds well to the
physicist's concept of the $\delta$ interaction.

Since $\Gamma$ is compact by assumption, the essential spectrum of
$H_{\alpha,\Gamma}$ equals $[0,\infty)$; our aim is to investigate
the asymptotic behavior of the negative eigenvalues as $\alpha\to
\infty$. It will be expressed in terms of the following comparison
operator,
%---------------%
\begin{equation} \label{compar}
S=-\Delta_{\Gamma} +K-M^2
\end{equation}
%---------------%
on $L^2(\Gamma,d\Gamma)$, where $\Delta_\Gamma = - g^{-1/2}
\partial_\mu g^{1/2} g^{\mu\nu} \partial_\nu$ is the
Laplace-Beltrami operator on $\Gamma$. We denote the $j$-th
eigenvalue of $S$ as $\mu_j$. Notice that it is bounded from above
by the $j$-th eigenvalue of $\Delta_\Gamma$ because the effective
potential
%---------------%
$$ K-M^2= -\frac{1}{4}(k_{+}-k_{-})^{2}\le 0\,; $$
%---------------%
the two coincide when $\Gamma$ is a sphere. Our first result then
reads as follows.

%---------------%
\begin{theorem} \label{t:comp}
(a) $\#\sigma_\mathrm{d}(H_{\alpha,\Gamma})\ge j$ holds for a
fixed integer $j$ if $\alpha$ is large enough. The $j$-th
eigenvalue $\lambda_j(\alpha)$ of $H_{\alpha,\Gamma}$ has then an
expansion of the form
%---------------%
\begin{equation} \label{asympt}
\lambda_j(\alpha) = -\frac{1}{4}\alpha^2 +\mu_j +\mathcal{O}(
\alpha^{-1} \ln\alpha) \quad \mathit{as} \quad \alpha\to\infty\,.
\end{equation}
%---------------%
(b) The counting function $\alpha\mapsto \#\sigma_\mathrm{d}
(H_{\alpha,\Gamma})$ behaves asymptotically as
%---------------%
\begin{equation} \label{count}
\#\sigma_\mathrm{d} (H_{\alpha,\Gamma}) = \frac{|\Gamma|}{16\pi}
\alpha^2 +\mathcal{O}( \alpha) \quad \mathit{for} \quad
\alpha\to\infty\,,
\end{equation}
%---------------%
where $|\Gamma|$ is the Riemann area of the surface $\Gamma$.
\end{theorem}
%---------------%

\subsection{Proof of Theorem~\ref{t:comp}} \label{pf_comp}

First we construct a family of layer neighborhoods of $\Gamma$.
Let $\{n(x):\, x\in\Gamma\}$ be a field of unit vectors normal to
the manifold. Such a field exists globally because $\Gamma$ is
orientable. Define a map $\mathcal{L}:\: \Gamma\times\mathbb{R}
\to \mathbb{R}^3$ by $\mathcal{L}(x,u)= x + un(x)$. Since $\Gamma$
is smooth by assumption, it is easy to see that there is an
$a_1>0$ such that for each $a\in(0,a_1)$ the restriction
%---------------%
\begin{equation} \label{neighb}
\mathcal{L}_a(x,u)= x + un(x),\quad (x,u)\in \mathcal{N}_a:=
\Gamma\times(-a,a)\,,
\end{equation}
%---------------%
is a diffeomorphism of $\mathcal{N}_a$ onto its image
$\Omega_a= \{ x\in\mathbb{R}^3:\: \mathrm{dist}(x, \Gamma)<a\}$.

We fix $a\in(0,a_1)$ and estimate (the negative spectrum of)
$H_{\alpha,\Gamma}$ using operators acting in the layer
$\Omega_a$. To this aim we define the quadratic forms $\eta
_{\alpha,\Gamma}^{\pm}[\cdot]$ with the domains $D(\eta
_{\alpha,\Gamma}^{+})=W_{0}^{2,1} (\Omega_a)$ and $D(\eta
_{\alpha, \Gamma}^{-})=W^{2,1}(\Omega_a)$, respectively, which
associate with a vector $\psi$ the value
%---------------%
%\begin{equation}
$$ \left\| \nabla \psi (x)\right\| _{L^{2}(\Omega_a)}^{2}
-\alpha \int_{\mathbb{R}^{3} }\left| \psi (x)\right|
^{2}d\mu_\Gamma(x)\,.$$
%\end{equation}
%---------------%
Both the forms are closed and bounded from below; we call the
self-adjoint operators in $L^{2}(\Omega_a)$ associated with them
$H_{\alpha,\Gamma }^{\pm}$. With this notation we can employ the
Dirichlet-Neumann bracketing argument \cite{RS} which yields the
bounds\footnote{This is the conventional way of expressing the
argument. A purist might object against inequalities between
operators having different domains. However, they make sense in
combination with the quadratic form version of the minimax
principle which is what we really need.}
%---------------%
\begin{equation} \label{bracke}
-\Delta _{\Sigma_{a}}^{N}\oplus H_{\alpha,\Gamma }^{-}\leq H_{\alpha,
\Gamma}\leq - \Delta _{\Sigma_{a}}^{D}\oplus H_{\alpha, \Gamma }^{+}\,,
\quad
\Sigma_{a}:= \mathbb{R}^{3} \setminus \overline{\Omega_a}\,.
\end{equation}
%---------------%
In the estimation operators the sets $\Omega_a$ and $\Sigma_{a}$
are decoupled, so $\sigma(H_{\alpha,\Gamma }^{\pm})$ is the union
of the two spectra. As long as we are interested in the negative
eigenvalues, we may take into account $H_{\alpha, \Gamma}^{\pm}$
only, because the ``exterior'' operators $\Delta _{\Sigma
_{d}}^{D}$ and $\Delta_{\Sigma _{d}}^{N}$ are positive by
definition.

In the next step we make use of the natural curvilinear
coordinates in $\Omega_a$. More specifically, we transform
$H_{\alpha, \Gamma}^{\pm}$ by means of the unitary operator
%---------------%
%\begin{equation}
$$
\hat U\psi =\psi \circ \mathcal{L}_a :\:
L^{2}(\Omega_a)\to L^{2} (\mathcal{N}_a ,d\Omega )\,.
%\end{equation}
$$
%---------------%
The measure $d\Omega$ is associated to the pull-back to
$\mathcal{N}_a$ of the Euclidean metric tensor in $\Omega_a$. We
denote this pull-back metric tensor by $G_{ij}$; it has the form
%---------------%
%\begin{equation}
$$
G_{ij}=
\left( \begin{array}{cc} (G_{\mu \nu }) & 0 \\
0 & 1 \end{array} \right)
,\quad G_{\mu \nu }=(\delta_{\mu}^{\sigma }-uh_{\mu
}\,^{\sigma }) (\delta_{\sigma}^{\rho}-uh_{\sigma
}\,^{\rho})g_{\rho\nu }\,,
%\end{equation}
$$
%--------------%
which yields $d\Omega := G^{1/2}d^{2}s\,du$ in local coordinates
with $ G:=\det (G_{ij})$ given by
%--------------%
%\begin{equation} \label{detGij}
$$
G=g\left[ (1-uk_{+})(1-uk_{-})\right] ^{2}=g(1-2Mu+Ku^{2})^{2}.
$$
%\end{equation}
%--------------%
Let $(\cdot ,\cdot )_{G}$ denote the inner product in the space
$L^{2}(\mathcal{N}_a,d\Omega )$. Then the operators $\hat
H_{\alpha,\Gamma}^{\pm}:= \hat UH_{\alpha,\Gamma}^{\pm}\hat
U^{-1}$ in $L^{2}(\mathcal{N}_a,d\Omega )$ are associated with the
forms $\psi \mapsto \eta _{\alpha,\Gamma}^{\pm} [\hat
U^{-1}\psi]$,
%---------------%
\begin{equation} \label{foret1}
\eta _{\alpha,\Gamma }^{\pm}[\hat U^{-1}\psi ]=
(\partial _{i}\psi ,G^{ij}\partial _{j}\psi )_{G}-\alpha
\int_{\Gamma }\left| \psi (s,0)\right| ^{2}d\Gamma\,,
\end{equation}
%--------------%
and they differ by their domains, $W_{0}^{2,1}(\mathcal{N}_a,
d\Omega )$ and $W^{2,1}(\mathcal{N}_a, d\Omega )$ for the $\pm$
sign, respectively. As above the expression $\psi (s,0)$ in
(\ref{foret1}) can be given natural meaning using the trace
mapping from $W_{0}^{2,1}(\mathcal{N}_a,d\Omega )$ or
$W^{2,1}(\mathcal{N}_a,d\Omega )$ to $L^{2}(\Gamma ,d\Gamma )$.

It is also useful to remove the factor $1-2Mu+Ku^2$ from the
weight $G^{1/2}$ in the inner product of
$L^{2}(\mathcal{N}_a,d\Omega)$. This is achieved by means of
another unitary transformation, namely
% ------------- %
\begin{equation} \label{operaB}
%$$
U\psi =(1-2Mu+Ku^2)^{1/2} \psi:\:L^{2}(\mathcal{N}_a,d\Omega)\to
L^{2} (\mathcal{N}_a,d\Gamma du)\,.
%$$
\end{equation}
% ------------- %
We will denote the inner product in $L^{2} (\mathcal{N}_a,d\Gamma
du)$ by $(\cdot,\cdot)_{g}$. The operators
$B^{\pm}_{\alpha,\Gamma}:=U \hat{H}^{\pm}_{\alpha,\Gamma}U^{-1}$
acting in $ L^{2}(\mathcal{N}_a,d\Gamma du)$ are associated with
the forms $b^{\pm}_{\alpha,\Gamma}$ given by
$b^{\pm}_{\alpha,\Gamma}[\psi] :=\eta^{\pm}_{\alpha,\Gamma}[(U\hat
U)^{-1}\psi]$ which again differ by their domains. A
straightforward computation, analogous to that performed in
\cite{DEK}, yields
% ------------- %
\begin{equation} \label{form1}
\begin{split}
b^{+}_{\alpha,\Gamma}[\psi] =\,&
(\partial_{\mu}\psi,G^{\mu\nu}\partial_{\nu}\psi)_{g}
+(\psi,(V_{1}+V_{2})\psi)_{g}+\|\partial_u\psi \|_{g}^{2}-\alpha
\int _{\Gamma } |\psi (s,0)|^{2}d\Gamma\,, \\
b^{-}_{\alpha,\Gamma}[\psi] =\,&
b^{+}_{\alpha,\Gamma}[\psi]+\int_{\Gamma } M_a(s)
|\psi(s,a)|^{2}d\Gamma -\int_{\Gamma } M_{-a}(s)|\psi(s,-a)|^{2}
d\Gamma
\end{split}
\end{equation}
% ------------- %
for $\psi$ from $W^{2,1}_{0}(\Omega_a,d\Gamma du)$ and
$W^{2,1}(\Omega_a, d\Gamma du)$, respectively. The quantity $M_u:=
(M-Ku)(1-2Mu+Ku^2)^{-1}$ here is the mean curvature of the
parallel surface characterized by a fixed value of $u$, and
% ------------- %
\begin{equation} \label{poteV1}
V_{1}:=g^{-1/2}(g^{1/2}G^{\mu \nu}J_{,\nu})_{,\mu }+ J
_{,\mu}G^{\mu \nu}J_{,\nu}\,,\quad
V_{2}:=\frac{K-M^{2}}{(1-2Mu+Ku^2)^2}
\end{equation}
% ------------- %
with $J:= \frac{1}{2}\ln (1-2Mu+Ku^2)$ is the effective
curvature-induced potential \cite{DEK}.

The operators $B^{\pm}_{\alpha,\Gamma}$ associated with the forms
(\ref{form1}) are still not easy to handle because the surface and
transverse variables are not decoupled. To get a rougher, but
still sufficient, estimate we notice that $1-2Mu+Ku^2$ can be
squeezed between the numbers $C_\pm(a):= (1\pm a\varrho^{-1})^2$,
where $\varrho:= \max (\{\left\| k_{+}\right\|_\infty, \left\|
k_{-}\right\|_\infty \})^{-1}$. Consequently, the matrix
inequality $C_-(a)g_{\mu\nu} \le G_{\mu\nu} \le C_+(a)g_{\mu\nu}$
is valid. Moreover, the first component of the effective potential
behaves as $\mathcal{O}(a)$ for $a\to 0$. Hence we have $|V_1|\le
va$ for some $v>0$, while $V_2$ can be squeezed between the
functions $C_{\pm}^{-2}(a) (K-M^2)$, both uniformly in the surface
variables. These observations motivate us to define the estimation
operators in the following way,
% ------------- %
\begin{equation} \label{crudeest}
\tilde{B}^{\pm}_{\alpha,a}:=S^{\pm}_a \otimes I + I\otimes
T^{\pm}_{\alpha,a}
\end{equation}
% ------------- %
with
% ------------- %
$$ S_a^{\pm} :=
-C_{\pm}(a)\Delta_{\Gamma } +C_{\pm}^{-2}(a)(K-M^2)\pm va
$$
% ------------- %
in $L^2(\Gamma,d\Gamma )\otimes L^2(-a,a)$, where
$T^{\pm}_{\alpha,a}$ are associated with the quadratic forms
% ------------- %
\begin{equation} \label{crudetrans}
\begin{split}
t^{+}_{\alpha,a}[\psi] &:=\, \int_{-a}^{a}|\partial_u\psi|^2 du -
\alpha |\psi (0)|^{2}\,, \\ t^{-}_{\alpha,a}[\psi] &:=\,
\int_{-a}^{a}|\partial_u\psi |^2 du - \alpha |\psi(0)|^2
-c_a(|\psi(a)|^{2}+|\psi(-a)|^2)\,.
\end{split}
\end{equation}
% ------------- %
In these relations $\psi$ belongs to $W^{2,1}_{0}(-a,a)$ and
$W^{2,1}(-a,a)$, respectively. In distinction to (\ref{form1}) the
coefficient $c_a:=2(\left\| M\right\|_{\infty }+\left\| K\right\|
_{\infty }a)$ in the boundary term of the second expression is
independent of the surface variables $s$. The operators \
(\ref{crudeest}) provide us with the sought estimate in view of
the obvious inequalities
% ------------- %
\begin{equation} \label{estimB}
\pm B_{\alpha,\Gamma}^{\pm}\leq \pm\tilde{B}_{\alpha,a}^{\pm}\,.
\end{equation}
% ------------- %
Since $\tilde{B}_{\alpha,a}^{\pm}$ have separated variables
their spectra express through those of their constituent
operators. To deal with the transverse part, we employ a simple
estimate the proof of which can be found in \cite{EY1}.
% ------------- %
\begin{lemma} \label{l:trans}
There are positive numbers $c,\: c_N$ such that each one of the
operators $T_{\alpha,a}^{\pm}$ has a single negative eigenvalue
$\kappa_{\alpha,a}^{\pm}$ satisfying the inequalities
%--------------%
%\begin{equation}
$$ -\frac{\alpha ^{2}}{4} \left(1+c_N \mathrm{e}^{-\alpha a/2}
\right) <\kappa_{\alpha,a}^{-} < -\frac{\alpha ^{2}}{4}<
\kappa_{\alpha,a}^{+}< -\frac{\alpha ^{2}}{4} \left(1 -8
\mathrm{e}^{-\alpha a/2} \right) $$
%\end{equation}
%--------------%
when the attraction is strong enough, $\alpha >c\,\max
\{a^{-1},c_a\}.$
\end{lemma}
% ------------- %

On the other hand, the surface part requires the
following result.

%----------------%
\begin{lemma}  \label{l:long}
The $j$-th eigenvalues of the operators $S^{\pm}_a$ satisfy the
asymptotic bounds $|\mu_{j,a}^{\pm} -\mu_j| \le m_j^{\pm}a$ with
some positive $m_j^{\pm}$ for all $a$ small enough.
\end{lemma}
%----------------%
\begin{proof}
We assume $a<\varrho$ so $C_-(a)$ is positive. Using the
definitions of $S^{\pm}_a$ and $C_{\pm}(a)$ we get easily the
asymptotic bound
%----------------%
%\begin{equation} \label{norU+C}
$$ \|S_a^{\pm}-C_{\pm}(a)S\|\leq \left( v+( \left\|
K\right\|_{\infty} +\left\| M\right\|_{\infty}^2)
\varrho^{-1}\right)a +\mathcal{O}(a^2) := m(a)\,. $$
%\end{equation}
%-----------------%
Combing this inequality with the minimax principle we find that
$|\mu_{j,a}^{\pm}-C_{\pm}(a)\mu_j|$ does not exceed $m(a)$. Using
once more the definition of $C_{\pm}(a)$ we conclude that
%-----------------%
$$ |\mu_{j,a}^{\pm}-\mu_j|\leq m(a)+a\left|(2\varrho^{-1}
+\varrho^{-2} a)\mu_j\right|\,, $$
%-----------------%
which implies the sought result for small $a$.
\end{proof}

Armed with these prerequisites we can now prove the asymptotic
expansion for eigenvalues of $H_{\alpha,\Gamma}$. By minimax
principle they are squeezed between the respective negative
eigenvalues of $\tilde{B}_{\alpha,a}^{\pm}$. Since each of the
operators $T_{\alpha,a}^{\pm}$ has a single negative eigenvalue,
the latter are of the form $\kappa_{\alpha,a}^{\pm}
+\mu_{j,a}^{\pm}$ provided $a$ is small and $\alpha$ is large
enough. For definiteness we suppose that these eigenvalues are
ordered in the same way as the $\mu_{j,a}^{\pm}$'s are. Choosing
%----------------%
\begin{equation} \label{dalpha}
a=a(\alpha):=6\alpha ^{-1}\ln \alpha
\end{equation}
%----------------%
and making use of the above two lemmata we find that
%----------------%
\begin{equation} \label{uplow}
\kappa _{\alpha,a}^{\pm} +\mu _{j,a}^{\pm}= -\frac{1}{4} \alpha ^2
+\mu_j +\mathcal{O}(\alpha^{-1}\ln\alpha )
\end{equation}
%----------------%
holds in this case as $a\to 0$. Since $\Gamma$ is compact the
spectrum of $S$ is purely discrete accumulating at infinity only.
Hence, to any positive integer $j$ there is an $\alpha_j$ such
that $\kappa _{\alpha,a}^+ +\mu_{j,a}^+ <0$ holds for
$\alpha>\alpha_j$. Consequently, $\tilde{B}_{\alpha,a}^+$ has at
least $j$ negative eigenvalues and the same is, of course, true
for $H_{\alpha,\Gamma}$. Furthermore, since the upper and lower
bound (\ref{uplow}) differ by the error term only, we arrive at
the claim (a).

Using minimax principle again we infer that there is a two-sided
estimate
%--------------%
\begin{equation} \label{numberest}
\#\sigma_\mathrm{d} (S^+_a) = \#\sigma_\mathrm{d}
(\tilde{B}_{\alpha, a}^+) \le \#\sigma_\mathrm{d}
(H_{\alpha,\Gamma}) \le \#\sigma_\mathrm{d} (\tilde{B}_{\alpha,
a}^-) = \#\sigma_\mathrm{d} (S^-_a)\,.
\end{equation}
%--------------%
Using (\ref{crudeest}) together with the definition of
$C_{\pm}(a)$ and the fact that the effective potential is bounded
we find that $\#\sigma_\mathrm{d} (S^{\pm}_a) =
\#\sigma_\mathrm{d} (S) (1+\mathcal{O}(a))$. Similarly, the
counting function for the operator (\ref{compar}) coincides with
that of $-\Delta_{\Gamma}$, up to the same error. Thus it suffices
to employ the well-known Weyl formula -- see, e.g., \cite{Ch2} --
to get the claim (b) and to conclude thus the proof.

\subsection{Remarks} \label{s:rems}

The assumption that the surface $\Gamma$ is connected was made
mostly for the sake of simplicity. The argument leading to the
asymptotic formula (\ref{asympt}) modifies easily to the case when
$\Gamma$ is a finite disjoint union of $C^4$ smooth compact
Riemann surfaces of finite genera. The situation is, of course,
substantially more complicated if the number of compact connected
components is infinite; in the next section we will discuss the
particular case when such a $\Gamma$ is periodic.

Furthermore, we have supposed that $\Gamma$ is a manifold without
a boundary. This was important in deriving the asymptotic
expansion (\ref{asympt}) because otherwise the eigenvalues of
$\mu_j$ of the comparison operator would not be properly defined.
On the other hand, the formula (\ref{count}) remains valid even if
$\Gamma$ has a nonempty and smooth boundary. It can be seen by an
easy modification of the above argument. If we construct the
neighborhood $\Omega_a$ in the described way, it will have the
boundary consisting of two parts. One of them,
$\partial\Omega_a^{(1)}$, contains as before points having normal
distance $a$ from $\Gamma$. The additional part,
$\partial\Omega_a^{(2)}$, is a subset of the normal surface to
$\Gamma$ at $\partial\Gamma$. The form domains of the estimation
operators will be again $W_{0}^{2,1} (\Omega_a)$ and
$W^{2,1}(\Omega_a)$, respectively. Consequently, the operators
$H^{\pm}_{\alpha,\Gamma}$ will satisfy Dirichlet and Neumann
conditions at the whole boundary. In particular, they will satisfy
these condition at the additional part $\partial\Omega_a^{(2)}$ of
the boundary, and the same will be true for the boundary
conditions which $S^{\pm}_a$ must satisfy at $\partial\Gamma$. The
eigenvalues of the last named operators no longer differ by an
$\mathcal{O}(a)$ term only. However, the Weyl asymptotics entering
(\ref{numberest}) is the same in both cases, because the
difference in the number of surface eigenvalues is hidden in the
error term.

It is also instructive to compare the formula (\ref{count}) with
the known estimates on the number of eigenvalues for generalized
Schr\"odinger operator with measure-type potentials such as the
modified Birman-Schwinger bound given in \cite[Sec.~4]{BEKS}. Our
result is valid in the asymptotic regime of strong coupling only
but by its very nature it has the correct semiclassical behavior.
On the other hand, the mentioned bound holds for any $\alpha>0$
but solvable examples, for instance with $\Gamma$ being a sphere
\cite{AGS, BEKS}, show that it may be rather crude for
$\#\sigma_\mathrm{d} (H_{\alpha, \Gamma})>1$.

%%%%%%%%%%%%%%%%%%%%%%%%%%%%%%%%%%%%%%%%%%%%%%%%%%%%%%%%%%%%%%%%%%

\section{Periodic surfaces}

\subsection{Floquet decomposition}

Let $\mathcal{T}\equiv \mathcal{T}_r(b)$ be a discrete Abelian
group of translations of $\mathbb{R}^3$ generated by an $r$-tuple
$\{b_i\}$ of linearly independent vectors, where $r=1,2,3$. The
starting point for the decomposition is a basic period cell
$\mathcal{C}$ of $\mathbb{R}^3$, which is a simply connected set
such that $\mathcal{C}_n:=  \mathcal{C}+ \sum_i n_i b_i$ is
disjoint with $\mathcal{C}$ for any $n=\{n_i\} \in\mathbb{Z}^r$
different from zero and $\bigcup_{n \in\mathbb{Z}^r} \mathcal{C}_n
= \mathbb{R}^3$. It is precompact if and only if $r=3$.
The simplest choice of such a period cell is
%----------------%
\begin{equation} \label{cell}
\mathcal{C} = \left\lbrace\, \sum_{i=1}^r t_ib_i\,:\: 0\le t_i<1
\right\rbrace \times \{b_i\}^\perp.
\end{equation}
%----------------%
The main geometric object of this section will be a $C^4$ smooth
Riemann surface $\Gamma \subset \mathbb{R}^3$, not necessarily
connected, which is supposed to be periodic, i.e. such that
$\mathcal{T}$ acts isometrically on $\Gamma$ and the quotient
space $\Gamma/\mathcal{T}$ is compact. A basic period cell of
$\Gamma$ is defined generally in terms of the group $\mathcal{T}$
and its fundamental domain \cite{Ch1}. In general the
decomposition of $\Gamma$ into period cells is independent of the
above decomposition of the Euclidean space. However, for our
purpose it is important that the two are consistent. Hence we
choose the period cell of $\Gamma$ in the form
$\Gamma_\mathcal{C}:= \Gamma\cap\mathcal{C}$. It is clear that
$\partial \Gamma_\mathcal{C} = \Gamma\cap
\partial\mathcal{C}$ is generally nonempty, in particular, if
$\Gamma$ is connected. The boundary is piecewise smooth if
$\partial\mathcal{C}$ has the same property.

We are interested again in the generalized Schr\"odinger operator
$H_{\alpha,\Gamma}$ with a $\delta$ interaction supported now by
the periodic surface. It is defined as above by means of the
quadratic form (\ref{form}); recall that Theorem~4.2 of
\cite{BEKS} used there does not require the compactness of
$\Gamma$. As usual in a periodic situation our main tool will be
the Floquet analysis. We introduce the family of quadratic forms
%--------------%
\begin{equation} \label{form_p}
\begin{split}
q_{\alpha,\theta}\left[ \psi \right] &= \| \nabla
\psi\|^2_{L^2(\mathcal{C})} -\alpha \int_{\mathcal{C} }
|\psi(x)|^{2} d\mu_\Gamma(x)\,, \\ \mathrm{Dom}(q_{\alpha,\theta})
&= \{\, \psi \in W^{2,1}(\mathcal{C})\,:\: \psi(x+b_i) =
\mathrm{e}^{i\theta_i} \psi(x) \,\} \,,
\end{split}
\end{equation}
%--------------%
where $\theta= \{\theta_i\} \in [0,2\pi)^r$, and denote by
$H_{\alpha,\theta}$ the self-adjoint operators associated with
them. For simplicity we will not indicate the dependence of these
forms and operators on $\Gamma$. Modifying the standard reasoning
\cite{RS, EY2} to the present situation we get the sought
decomposition.
%---------------%
\begin{lemma} \label{l:Floq}
There is a unitary map $\mathcal{U}: \, L^2(\mathbb{R}^3) \to
\int^\oplus_{[0,2\pi)^r} L^2(\mathcal{C})\, d\theta$ such that
%----------------%
\begin{equation} \label{FloqHam}
\mathcal{U} H_{\alpha,\Gamma} \mathcal{U}^{-1} =
\int^\oplus_{[0,2\pi)^r} H_{\alpha,\theta}\, d\theta
\end{equation}
%----------------%
and
%----------------%
\begin{equation} \label{Floqspec}
\sigma(H_{\alpha,\Gamma}) = \bigcup_{[0,2\pi)^r}
\sigma(H_{\alpha,\theta})\,.
\end{equation}
%----------------%
The spectrum of $H_{\alpha,\theta}$ is purely discrete if $\,r=3$
while $\sigma_\mathrm{ess}(H_{\alpha,\theta})= [0,\infty)$ if
$\,r=1,2$; the eigenvalues (conventionally arranged in the
ascending order, with their multiplicity taken into account) are
continuous functions of the quasimomenta $\theta_i$.
\end{lemma}
%---------------%

Consequently, behavior of the spectral bands of
$H_{\alpha,\Gamma}$ can be found through properties of eigenvalues
of the fiber operators. The difference between the situation with
$r=3$ and the ``partially periodic'' cases, $r=1,2$, is that in
the former we will get the asymptotic behavior for all bands. Of
course, the error term will not be uniform in the band index.
Another difference is that in the case $r=3$ the spectrum is known
to be absolutely continuous \cite{SS}, even under weaker
assumptions than used here, while for $r=1,2$ this remains to be
an open problem.

\subsection{Fiber operator eigenvalues asymptotics} \label{s:fiber}

As before we need a comparison operator. In the present case it is
defined on $L^2(\mathcal{C},\mu_\Gamma)\cong
L^2(\Gamma_\mathcal{C}, d\Gamma)$ by
%---------------%
\begin{equation} \label{compar_per}
S_\theta = -\Delta_{\Gamma} +K-M^2
\end{equation}
%---------------%
with the domain consisting of those $\phi \in
W^{2,1}(\Gamma_\mathcal{C})$ with $\Delta_{\Gamma}\phi\in
L^2(\Gamma_\mathcal{C}, d\Gamma)$. If $\Gamma_\mathcal{C}$ has a
nontrivial boundary we have to require in addition that $\phi$
satisfies the Floquet conditions at the points of $\partial
\Gamma_\mathcal{C}$. One can always choose the atlas in such a way
that the local charts are periodic with respect to the group
$\mathcal{T}$. In that case the conditions read $\phi(x+b_i) =
\mathrm{e}^{i\theta_i} \phi(x)$ and
%---------------%
\begin{equation} \label{der_per}
\frac{\partial\phi(x+b_i)}{\partial s_\mu} =
\mathrm{e}^{i\theta_i} \frac{\partial\phi(x)}{\partial s_\mu}\,,
\quad 1\le i\le r\,, \quad \mu=1,2\,,
\end{equation}
%---------------%
for derivatives with respect to the surface
coordinates\footnote{One can also use a coordinate-free way, for
instance, by bending the elementary cell into a torus and moving
the quasimomentum from the boundary conditions into the
operator.}. Since $\Gamma_\mathcal{C}$ is precompact and the
curvatures involved are bounded, the spectrum of $S_\theta$ is
purely discrete for each $\theta \in [0,2\pi)^r$; we denote the
$j$-th eigenvalue of $S_\theta$ as $\mu_j(\theta)$.

%---------------%
\begin{theorem} \label{t:comp_per}
Under the stated assumptions the following claims are valid:
\\[1mm]
%---------------%
(a) Fix $\lambda$ as an arbitrary number if $r=3$ and a
non-positive one for $r=1,2$. To any $j\in\mathbb{N}$ there is
$\alpha_j>0$ such that $H_{\alpha,\theta}$ has at least $j$
eigenvalues below $\lambda$ for any $\alpha>\alpha_j$ and
$\theta \in [0,2\pi)^r$. The
$j$-th eigenvalue $\lambda_j(\alpha, \theta)$ has then the
expansion
%---------------%
\begin{equation} \label{asympt_per}
\lambda_j(\alpha, \theta) = -\frac{1}{4}\alpha^2 +\mu_j(\theta)
+\mathcal{O}( \alpha^{-1} \ln\alpha) \quad \mathit{as} \quad
\alpha\to\infty\,,
\end{equation}
%---------------%
where the error term is uniform with respect to $\theta$. \\[1mm]
%---------------%
(b) If the set $\sigma(S):= \bigcup_{\theta \in
[0,2\pi)^r}\sigma(S_\theta)$ has a gap separating a pair of bands,
then the same is true
for $\sigma (H_{\alpha,\Gamma})$ if $\alpha$ is large enough.
\end{theorem}
%---------------%
\begin{proof}
The argument is the same as in Section~\ref{pf_comp}, one has just
to modify the domains of the quadratic forms involved and to check
that the used estimates are uniform in $\theta$ which follows from
the continuity of the Floquet eigenvalues. \end{proof}

\subsection{Compactly disconnected periodic surfaces}

By the second part of Theorem~\ref{t:comp_per} the operator
$H_{\alpha,\Gamma}$ has open spectral gaps in the asymptotic
regime if the comparison operator has the same property. The
latter may or may not be true depending on the geometry of
$\Gamma$. The situation is different, however, if $\Gamma$ is not
connected and each one of its connected component is compact and
contained in (an interior of) a translate of the period cell
$\mathcal{C}$. Let us stress that the last named property is a
nontrivial assumption; to see that this is the case imagine a
family of annular surfaces interlaced neighborwise into an
infinite periodic ``chain''\footnote{In fact the proof of
Theorem~\ref{t:comp_per_comp} can be modified to this case too. We
use this assumption to avoid a cumbersome formulation needed in
more general situations.}. An equally important observation is
that while in Section~\ref{s:fiber} the choice of the basic period
cell $\mathcal{C}$ was mostly irrelevant and one could settle for
the simplest one represented by (\ref{cell}), it clearly matters
here. To give an example\footnote{A proper name would be an
``Australian gift shop'' example.} consider an infinite array of a
boomerang-shaped surfaces: they cannot be stacked in individual
rectangular boxes if one wants them to be close enough to each
other.

If the assumptions of this section are valid then the domain of
the comparison operator is independent of the Floquet conditions
(\ref{der_per}). In that case, $S_\theta$ does not depend on
$\theta$, and it has the form of a finite direct sum  of operators
of the type (\ref{compar}) for finite-genae surfaces to which the
basic period cell $\Gamma_\mathcal{C}$, now automatically closed,
can be decomposed (we have noted that in the first remark of
Section~\ref{s:rems}). On the other hand, the Floquet
decomposition (\ref{FloqHam}) of the operator $H_{\alpha,\Gamma}$
is nontrivial and its spectral bands have generically nonzero
widths. By the absolute continuity result mentioned above, we know
that this is always true in the ``fully periodic'' case, $r=3$,
while for $r=1,2$ the analogous claim is presently just a
conjecture.

An easy way to estimate the spectral band widths is to employ a
bracketing argument again. Inspecting the domains of the quadratic
forms (\ref{form_p}) we see that the Floquet eigenvalues can be
bound from above and below if the boundary conditions for the
fiber operator $H_{\alpha,\theta}$ are changed to Dirichlet and
Neumann, respectively. Since $\mathrm{dist}(\partial\mathcal{C},
\Gamma_\mathcal{C})>0$ holds by assumption, the neighborhood
$\Omega_a$ of $\Gamma_\mathcal{C}$ is contained in the interior of
$\mathcal{C}$ for all $a>0$ small enough. The negative part of the
spectrum of the two estimation operators can be then treated in
the exactly the same way as in the case of a compact surface,
singly or finitely connected, because the ``exterior'' region
$\Gamma_\mathcal{C} \setminus \mathcal{N}(a)$ contributes to the
positive part only (more exactly, from the first eigenvalue up if
$r=3$). We arrive thus at the following result.

%---------------%
\begin{theorem} \label{t:comp_per_comp}
Let $\Gamma$ be a $C^4$ smooth periodic surface such that each one
of its connected component is compact and contained in (an
interior of) a translate of a fixed period cell $\mathcal{C}$ of
the group $\mathcal{T}$. Denote by $\mu_j$ the $j$-th eigenvalue
of the comparison operator (\ref{compar_per}). Then the $j$-th
Floquet eigenvalue $\lambda_j(\alpha, \theta)$ from the
decomposition (\ref{FloqHam}) of the operator $H_{\alpha,\Gamma}$
behaves asymptotically as
%---------------%
\begin{equation} \label{asympt_per_comp}
\lambda_j(\alpha, \theta) = -\frac{1}{4}\alpha^2 +\mu_j
+\mathcal{O}( \alpha^{-1} \ln\alpha) \quad \mathit{for} \quad
\alpha\to\infty\,,
\end{equation}
%---------------%
with the error term uniform with respect to the quasimomenta
$\theta$. Consequently, the number of open gaps in
$\sigma(H_{\alpha, \Gamma})$ exceeds any fixed integer if $\alpha$
is large enough.
\end{theorem}
%---------------%

%%%%%%%%%%%%%%%%%%%%%%%%%%%%%%%%%%%%%%%%%%%%%%%%%%%%%%%%%%%%%%%%%%

\section{Concluding remarks}

\subsection{Curves in the plane}

A two-dimensional analogue of the present results was discussed in
\cite{EY1, EY2}, with $\Gamma$ being a smooth loop or an infinite
smooth connected periodic curve. We have derived asymptotic
expansions of the form (\ref{asympt}) and (\ref{asympt_per}) where
$\mu_j$ and $\mu_j(\theta)$, respectively, are eigenvalues of the
operator
%---------------%
\begin{equation} \label{compar_curve}
S = -\partial_s^2 -\frac{1}{4} k(s)^2\,.
\end{equation}
%---------------%
Here $s$ is the arc length variable and $k$ is the signed
curvature of $\Gamma$. The boundary conditions were periodic for
the loop and the Floquet ones over the period in the other case.
As in the three-dimensional situation, it is easy to extend these
results to Hamiltonians with the $\delta$ interaction supported by
a family of curves, be it a finite number of nonintersecting loops
or a periodic system with multiple curves\footnote{A model similar
to the last named case was treated by a different technique in
\cite{KK} and earlier in \cite{FK}. The setting used in these
papers differs slightly from the present one. It concerns the
roles of the coupling and spectral parameters which are switched
there.}.

The latter includes families of curves periodic in two directions,
i.e. $r=2$ in the terminology of the preceding section. In that
case we know from \cite{BSS} that the spectrum of
$H_{\alpha,\Gamma}$ is purely absolutely continuous so none of the
spectral bands is degenerate; for $r=1$ this is an open problem
again. From the viewpoint of open gaps, it is the absence of
noncompact connected components of $\Gamma$ which is important. If
$\Gamma$ can be broken into finite families of loops confined
within the interior of the period cells, the analogue of
Theorem~\ref{t:comp_per_comp} is valid and the system has many
gaps for large $\alpha$. By \cite{EY2} a single periodic connected
curve which is not a straight line gives rise to an open gap for
large $\alpha$, because the comparison operator
(\ref{compar_curve}) has the same property. It is not a priori
clear whether the same is true for two or more such curves,
because then we compare with a union of band spectra in which the
gaps in one component may overlap with bands in the other one. It
is not excluded, of course, that some gaps may survive since the
curves have the same periodicity group; the problem deserves a
deeper investigation.

\subsection{Semiclassical interpretation}

The results discussed here and in the earlier work mentioned in
the introduction can be viewed also from a different perspective.
Recall that the deviation of the spectrum of $H_{\alpha,\Gamma}$
from the one corresponding to the ideal manifold described by the
comparison operator (\ref{compar}) are due to quantum tunneling.
Hence they must be sensitive to the appropriate parameter, i.e.
the Planck's constant if we reintroduce it into the picture.
However, the operator $-h^2\Delta -v\delta(x-\Gamma)$ is the $h^2$
multiple of (\ref{formal}) if we denote $\alpha:= vh^{-2}$. In
this sense the obtained asymptotic formulae represent a
semiclassical approximation.

\subsection{Open problems}

One can ask whether the ``wide'' gaps which one has if $\alpha$ is
large and $\Gamma$ is decomposed into compact components will
persist when the assumption about non-connected character of
$\Gamma$ is weakened. It is natural to conjecture that the answer
depends on properties of the corresponding operator
(\ref{compar}). For the Laplace-Beltrami operator a construction
of connected periodic manifolds exhibiting gaps has been presented
recently \cite{Po1}. It is based on connecting compact components
of a non-connected surface by thin cylinders. It is worth
examining what will be the effect of the curvature-induced
potential $K-M^2$ which represents additional ``potential wells''
at the cylinders and connecting necks.

Spectral properties of the Laplace-Beltrami operator were studied
also for other surface classes such as locally perturbed periodic
ones for which eigenvalues in the gaps may appear \cite{Po2}. A
similar behavior may be expected for the operator (\ref{compar})
and one can ask whether the same will be true asymptotically for
the corresponding $H_{\alpha,\Gamma}$. Our present method can
yield information only on eigenvalues below the threshold of
$\sigma_\mathrm{ess}(H_{\alpha,\Gamma})$ because it employs the
minimax principle in a substantial way, so another approach is
needed.

An extension to higher dimensions, to an $m$-dimensional $\Gamma$
in $\mathbb{R}^n$, is also interesting. One conjectures that a
similar asymptotic formula will be valid with the effective
potential replaced by the function of the principal curvatures
derived in \cite{To}, see also \cite{FH}. It is needed, however,
that the operator corresponding to the symbol (\ref{formal}) makes
sense. If the $\mathrm{codim}\,\Gamma= n-m$ equals one, it is
defined as here in terms of quadratic forms. If $n-m=2,3$ one can
proceed as in \cite{EK2} using generalized boundary conditions
(and $-\alpha^2/4$ will be replaced by the point-interaction
eigenvalue in dimension $n-m$); for $\mathrm{codim}\,\Gamma>3$
there is no meaningful operator $H_{\alpha,\Gamma}$, at least as
long as we stay within the Hilbert-space theory. \\ [2mm]
Useful comments by S.~Kondej, D.~Krej\v{c}i\v{r}\'{\i}k, and
P.~Kuchment are gratefully acknowledged. I also appreciate the
referee who checked every sentence twice at least.

\bibliographystyle{amsalpha}

\begin{thebibliography}{A}

\bibitem[AGS]{AGS} J.-P.~Antoine, F.~Gesztesy, J.~Shabani:
\textit{Exactly solvable models of spere interaction in quantum
mechanics}, J. Phys. \textbf{A20} (1987), 3627--3712.

\bibitem[BS\v{S}]{BSS} M.S.~Birman, T.A.~Suslina, R.G.~Shterenberg:
\textit{Absolute continuity of the two-dimensional Schr\"odinger
operator with delta potential concentrated on a periodic system of
curves}, Algebra i Analiz \textbf{12} (2000), 140--177; translated
in St. Petersburg Math. J. \textbf{12} (2001), 535--567.

\bibitem[BEK\v{S}]{BEKS} J.F.~Brasche, P.~Exner, Yu.A.~Kuperin,
P.~\v{S}eba: \textit{Schr\"odinger operators with singular
interactions}, J. Math. Anal. Appl. \textbf{184} (1994), 112--139.

\bibitem [Ch1]{Ch1} I.~Chavel, \textit{Riemannian Geometry; A Modern
Perspective}, Cambridge University Press 1993.

\bibitem [Ch2]{Ch2} I.~Chavel, \textit{The Laplacian on Riemannian manifolds},
Proc. ICMS Instructional Conference ``Spectral Theory and
geometry'', Cambridge University Press 1999, pp. 30--75.

\bibitem[dC]{dC}
R.C.T. da Costa: \textit{Quantum mechanics of a constrained
particle}, Phys. Rev. \textbf{A23} (1981), 1982--1987.

\bibitem[DE]{DE}
P.~Duclos, P.~Exner: \textit{Curvature-induced bound states in
quantum waveguides in two and three dimensions}, Rev. Math. Phys.
\textbf{7} (1995), 73--102.

\bibitem[DEK]{DEK}
P.~Duclos, P.~Exner, D.~Krej\v{c}i\v{r}ik: \textit{Bound states in
curved quantum layers}, Commun. Math. Phys. \textbf{223} (2001),
13--28.

\bibitem[EI]{EI}
P.~Exner, T.~Ichinose: \textit{Geometrically induced spectrum in
curved leaky wires}, J. Phys. \textbf{A34} (2001), 1439--1450.

\bibitem [EK1]{EK} P.~Exner, S.~Kondej, \textit{Bound states due to a
strong $\delta$ interaction supported by a curved surface}, J.
Phys. \textbf{A36} (2003), 443--457.

\bibitem [EK2]{EK2}
P.~Exner, S.~Kondej: \textit{Curvature-induced bound states for a
$\delta$ interaction supported by a curve in $\mathbb{R}^3$}, Ann.
H.~Poincar\'{e} \textbf{3} (2002), 967--981.

\bibitem [EY1]{EY1} P.~Exner, K.~Yoshitomi, \textit{Asymptotics of
eigenvalues of the Schr\"odinger operator with a strong
$\delta$-interaction on a loop}, J. Geom. Phys. \textbf{41}
(2002), 344--358.

\bibitem [EY2]{EY2} P.~Exner, K.~Yoshitomi, \textit{Band gap of the
Schr\"odinger operator with a strong $\delta$-interaction on a
periodic curve}, Ann. H.~Poincar\'e \textbf{2} (2001), 1139--1158.

\bibitem [EY3]{EY3} P.~Exner, K.~Yoshitomi, \textit{Persistent currents
for 2D Schr\"odinger operator with a strong $\delta$-interaction
on a loop}, J. Phys. \textbf{A35} (2002), 3479--3487.

\bibitem[FK]{FK}
A.~Figotin, P.~Kuchment: \textit{Band-gap structure of spectra
of periodic dielectric and acoustic media I, II}, SIAM J. Appl.
Math. \textbf{56} (1996), 68--88, 1561--1620.

\bibitem[FH]{FH}
R.~Froese, I.~Herbst: \textit{Realizing holonomic constraints in
classical and quantum mechanics}, Commun. Math. Phys. \textbf{220}
(2001), 489--535.

\bibitem[JK]{JK}
H.~Jensen, H.~Koppe: \textit{Quantum mechanics with constraints},
Ann. Phys. \textbf{63} (1971), 586--591.

\bibitem [Kli]{Kli} W.~Kligenberg, \textit{A Course in Differential Geometry},
Springer Verlag, New York 1978.

\bibitem[KK]{KK}
P.~Kuchment, L.~Kunyansky: \textit{Spectral properties of high-contrast
band-gap materials and operators on graphs}, Experimental Math.
\textbf{8} (1998), 1--28.

\bibitem[KZ]{KZ}
P.~Kuchment, Hongbiao Zeng: \textit{Convergence of spectra of
mesoscopic systems collapsing onto a graph}, J. Math. Anal. Appl.
\textbf{258} (2001), 671--700.

\bibitem[Po1]{Po1} O.~Post:
\textit{Periodic manifolds with spectral gaps}, J. Diff. Eq., to
appear;  \texttt{math-ph/0207017}.

\bibitem[Po2]{Po2} O.~Post:
\textit{Eigenvalues in spectral gaps of a perturbed periodic
manifold}, \texttt{math-ph/0207018}.

\bibitem [RS]{RS} M.~Reed, B.~Simon, \textit{Methods of Modern Mathematical
Physics, IV.~Analysis of Operators}, Academic Press, New York
1978.

\bibitem[RuS]{RuS}
J.~Rubinstein, M.~Schatzman, \textit{Variational problems on
multiply connected thin strips, I.~Basic estimates and convergence
of the Laplacian spectrum}, Arch. Rat. Mech. Anal. \textbf{160}
(2001), 271--308.

\bibitem[S\v{S}]{SS} T.A.~Suslina, R.G.~Shterenberg:
\textit{Absolute continuity of the spectrum of the Schr\"odinger
operator with the potential concentrated on a periodic system of
hypersurfaces}, Algebra i Analiz \textbf{13} (2001), 197--240.

\bibitem[To]{To}
J.~Tolar: \textit{On a quantum mechanical d'Alembert principle},
in ``Group Theoretical Methods in Physics'', Lecture Notes in
Physics, vol.~313, Springer, Berlin 1988; pp.~268-274.

\end{thebibliography}

\end{document}